\documentclass[aps,pra,reprint,graphicx,superscriptaddress]{revtex4-1}

\usepackage{color}
\usepackage{amsmath}
\usepackage{graphicx}
\usepackage{multirow}
\usepackage{amsfonts}
\usepackage{amssymb}
\usepackage{amscd}
\usepackage{multirow}
\usepackage{dcolumn}
\usepackage{bm}
\usepackage{float}

\begin{document}

\title{Phase amplification in optical interferometry with weak measurement}

\author{Li Li}
\email{eidos@ustc.edu.cn}
\affiliation{Hefei National Laboratory for Physical Sciences at the Microscale and Department
of Modern Physics, University of Science and Technology of China, Hefei, Anhui 230026, China}
\affiliation{CAS Center for Excellence in Quantum Information
and Quantum Physics, University of Science and Technology of China, Hefei, Anhui 230026, China}

\author{Yuan Li}
\affiliation{Hefei National Laboratory for Physical Sciences at the Microscale and Department
of Modern Physics, University of Science and Technology of China, Hefei, Anhui 230026, China}
\affiliation{CAS Center for Excellence in Quantum Information
and Quantum Physics, University of Science and Technology of China, Hefei, Anhui 230026, China}

\author{You-Lang Zhang}
\affiliation{Hefei National Laboratory for Physical Sciences at the Microscale and Department
of Modern Physics, University of Science and Technology of China, Hefei, Anhui 230026, China}
\affiliation{CAS Center for Excellence in Quantum Information
and Quantum Physics, University of Science and Technology of China, Hefei, Anhui 230026, China}

\author{Sixia Yu}
\affiliation{Hefei National Laboratory for Physical Sciences at the Microscale and Department
of Modern Physics, University of Science and Technology of China, Hefei, Anhui 230026, China}
\affiliation{CAS Center for Excellence in Quantum Information
and Quantum Physics, University of Science and Technology of China, Hefei, Anhui 230026, China}

\author{Chao-Yang Lu}
\affiliation{Hefei National Laboratory for Physical Sciences at the Microscale and Department
of Modern Physics, University of Science and Technology of China, Hefei, Anhui 230026, China}
\affiliation{CAS Center for Excellence in Quantum Information
and Quantum Physics, University of Science and Technology of China, Hefei, Anhui 230026, China}

\author{Nai-Le Liu}
\affiliation{Hefei National Laboratory for Physical Sciences at the Microscale and Department
of Modern Physics, University of Science and Technology of China, Hefei, Anhui 230026, China}
\affiliation{CAS Center for Excellence in Quantum Information
and Quantum Physics, University of Science and Technology of China, Hefei, Anhui 230026, China}

\author{Jun Zhang}
\email{zhangjun@ustc.edu.cn}
\affiliation{Hefei National Laboratory for Physical Sciences at the Microscale and Department
of Modern Physics, University of Science and Technology of China, Hefei, Anhui 230026, China}
\affiliation{CAS Center for Excellence in Quantum Information
and Quantum Physics, University of Science and Technology of China, Hefei, Anhui 230026, China}

\author{Jian-Wei Pan}
\affiliation{Hefei National Laboratory for Physical Sciences at the Microscale and Department
of Modern Physics, University of Science and Technology of China, Hefei, Anhui 230026, China}
\affiliation{CAS Center for Excellence in Quantum Information
and Quantum Physics, University of Science and Technology of China, Hefei, Anhui 230026, China}

\date{\today}

\begin{abstract}
Improving the phase resolution of interferometry is crucial for
high-precision measurements of various physical quantities.
Systematic phase errors dominate the phase uncertainties in most realistic optical interferometers.
Here we propose and experimentally demonstrate a weak measurement scheme to considerably suppress the phase uncertainties
by the direct amplification of phase shift in optical interferometry. Given an initial ultra-small phase
shift between orthogonal polarization states, we observe the phase
amplification effect with a factor of 388. Our weak measurement scheme
provides a practical approach to significantly improve the interferometric
phase resolution, which is favorable for precision measurement applications.
\end{abstract}

\maketitle

\section{introduction}

Interferometry is one of the most important metrology tools by transforming
the measurements of various physical quantities into phase measurements. The
precision of measurement highly depends on the phase resolution of
interferometry. The phase resolution is limited by the uncertainty
contributions of two parts, i.e., statistical phase errors and systematic
phase errors. For a given sample size $N$, the statistical errors are
described as the standard quantum limit, i.e., $\frac{1}{\sqrt{N}}$.
Quantum-enhanced measurements as resource-efficient methods are often used
to beat the standard quantum limit~\cite%
{Giovannetti04,Giovannetti11,Boto00,Holland93,Caves81}. For instance, given
a $n$-qubit entangled $N00N$ state $\frac{1}{\sqrt{2}}(|0\rangle ^{\otimes {n%
}}+|1\rangle ^{\otimes {n}})$, due to its characteristic of phase
super-sensitivity~\cite{Boto00}, the statistical errors can be decreased
down to $\frac{1} {\sqrt{nN}}$.

In realistic optical interferometers, the systematic phase errors due to
device imperfections often dominate the phase uncertainty, which cannot be
reduced by averaging. For instance, the phase uncertainty of measuring the
phase shift between two orthogonal polarization components is mainly
attributed to alignment errors of polarization devices in interferometers.
In such case, the phase uncertainty is bounded by $\frac{1}{\sqrt{e}}$,
which $e$ is the polarization extinction ratio (PER) of polarization
devices, since the minimal probability of projection measurement is at the
order of $\frac{1}{e}$. Therefore, to improve the phase resolution of
optical interferometers reducing the contribution of the intrinsic phase
errors is crucial.

\begin{figure}[h]
\centering
\includegraphics[width=8 cm]{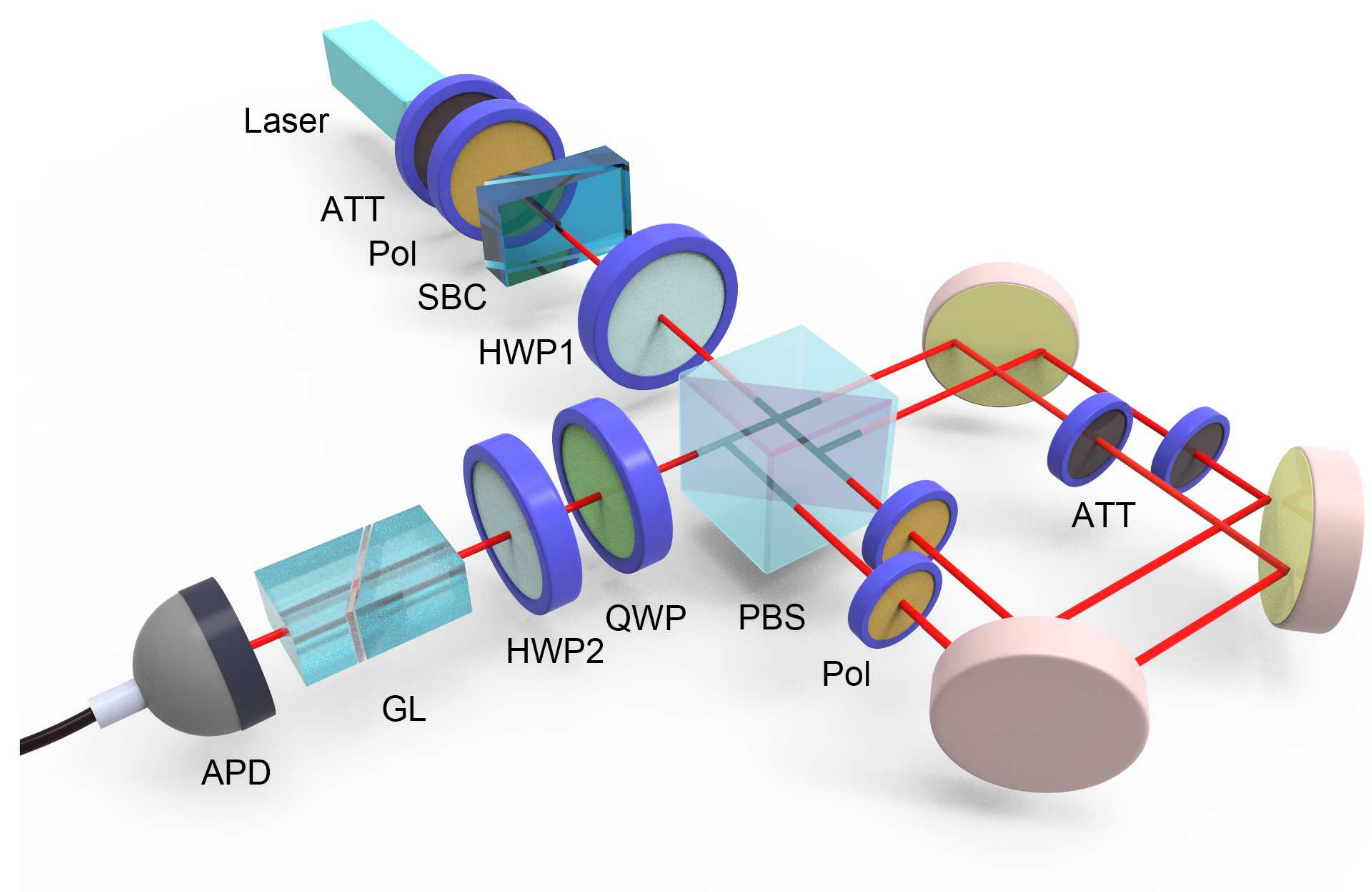}
\caption{Experimental setup. Focused laser pulses with a center wavelength
of 785 nm pass through a set of attenuators to prepare the incident photon
source, and the attenuated intensity of the source is less than one million
photons per second. A polarizer is placed after the attenuator to prepare
the initial polarization state of $\frac{1}{\protect\sqrt{2}}(|H\rangle
+|V\rangle )$. An initial tiny phase is generated by the SBC between $%
|H\rangle $ and $|V\rangle $. HWP1 is used to implement the unitary rotation
$U\left( \protect\alpha \right) $. Then the beam passes through a
Sagnac-like interferometer with two tunable attenuators and two polarizers
in two different paths. The angles of the polarizers are placed at $%
|H\rangle $ and $|V\rangle $ for the transmitted and reflected photons,
respectively, in order to increase the polarization visibility. The tunable
attenuators are used to balance the intensity in two paths. The two arms of
the interferometer are separate with only 4 mm so that the stability of
interferometer can be effectively guaranteed. A QWP set at the angle of $%
\protect\pi /4$ and HWP2 are used to project polarization states to $%
(|H\rangle +e^{i\protect\beta }|V\rangle )/\protect\sqrt{2}$, in which the
modulated phase delay $\protect\beta $ is implemented by tuning the angle of
HWP2. ATT: attenuator; Pol: polarizer; SBC: Soleil-Babinet Compensator; HWP:
half-wave plate; QWP: quarter-wave plate; PBS: polarizing beam splitter; GL:
Glan-Laser Calcite Polarizer; APD: avalanche photodiode.}
\label{fig1}
\end{figure}

We first analyze the interferometric phase uncertainty using single-qubit
states. Considering the typical configuration of a realistic optical
interferometer to measure a small initial phase, the detection probability
at one output port of the interferometer can be calculated as
\begin{equation}
p=\frac{1}{2}(1+\cos (\phi +\varphi )),  \label{p}
\end{equation}%
where $\phi $ is the phase to be measured, and $\varphi $ is the systematic
phase error due to device imperfections. Generally, $\varphi $ can be
described as a zero-mean random phase error following a Gaussian
distribution $e^{-\frac{\varphi ^{2}}{2\rho ^{2}}}$ with a standard
deviation of $\rho $. Given a sample size of $N$, the total phase
uncertainty is (see Eq. (A10) in Appendix A)
\begin{equation}
\Delta \phi =\sqrt{\rho ^{2}+\frac{1}{N}},  \label{pu1}
\end{equation}%
where the systematic phase errors and the statistical phase errors are
related with the terms of $\rho $ and $\frac{1}{\sqrt{N}}$, respectively,
and $\rho $ is often much larger than $\frac{1}{\sqrt{N}}$ for realistic
optical interferometers. When using $N00N$ states instead of single-qubit
states, the detection probability ($p^{(N)}$) is then changed to
\begin{equation}
p^{(N)}=\frac{1}{2}(1+\cos (n\phi +\varphi )). \label{pn}
\end{equation}%
The total phase uncertainty in this scenario is $\sqrt{\frac{\rho ^{2}}{n^2}+%
\frac{1}{nN}}$ (see Eq. (A13) in Appendix A). However, preparing multi-qubit entangled states with large $n$ is proven notoriously challenging. The largest number of entangled photons created so far is 10~\cite{Wang16, Chen17} in a probabilistic way, limiting the practical advantage of using $N00N$ states to improve the phase resolution.

Weak measurement is a new quantum-mechanical approach to further suppress
the phase uncertainty. The concept of weak measurement was discovered in
1980s~\cite{Aharonov88}. So far diverse weak measurement schemes have been
proposed for the applications of precise measurements due to the advantages
of amplification effects~\cite%
{Aharonov90,Feizpour11,Zilberberg11,Wu12,Strubi13,Pang14}, which
can be used to effectively overcome the device imperfections. Experiments
using weak measurement schemes have demonstrated to perform the precise
measurements of certain quantities such as ultra-small transverse split~\cite%
{Hosten08}, beam deflection~\cite{Dixon09}, light chirality~\cite%
{Gorodetski12} and angular rotation~\cite{Magana14}.
Specifically, Brunner and Simon~\cite{Brunner10} proposed a scheme to convert the quantity of time delay to frequency shift using imaginary weak-value amplification.
Based on this scheme, Xu \emph{et al.}~\cite{Xu13} demonstrated to measure a time delay at the order of attosecond, corresponding to a phase resolution of a few mrad,
using a commercial light-emitting diode and a spectrometer, and Salazar-Serrano \emph{et al.}~\cite{SS14} implemented a sub-pulse-width time delay measurement as small as 22 femtoseconds using
a femtosecond fiber laser. Recently, Qiu \emph{et al.}~\cite{Xiaodong17} reported an approach to convert phase estimation to intensity measurement with imaginary weak-value amplification, showing a phase resolution $\sim$ $1$ mrad.

\section{scheme}

In this Letter, we propose and experimentally demonstrate a new weak
measurement scheme, in which the amplification of phase shifts in an
interferometer can be directly implemented. The experimental results clearly
show that the phase resolution of interferometer can be improved with 2$\sim$%
3 orders of magnitudes besides the conventional techniques for phase
stabilization.

In our weak measurement scheme, it is assumed that the input qubit state is $%
\frac{1}{\sqrt{2}}(|H\rangle +e^{i\phi }|V\rangle)$, where $|H\rangle \left(
|V\rangle \right) $ represents the horizontal (vertical) polarization state
of photons, and $\phi$ is the initial phase shift to be measured. After
passing a half-wave plate with a unitary rotation
\begin{eqnarray}  \label{U}
U\left( \alpha \right) = \left(
\begin{array}{cc}
\cos \alpha & \sin \alpha \\
-\sin \alpha & \cos \alpha%
\end{array}%
\right),
\end{eqnarray}
where $\alpha =\pi /4-\varepsilon$ and $\varepsilon \ll 1,$ the input state
is transformed to $\left( \cos \alpha +e^{i\phi }\sin \alpha \right)
|H\rangle +\left( -\sin \alpha +e^{i\phi }\cos \alpha \right) |V\rangle$.
Then the final phase shift between $|H\rangle $ and $|V\rangle $ components
is%
\begin{equation}
\phi ^{\prime }=ang\left( \sin 2\varepsilon \cos \phi +i\sin \phi \right) ,
\label{pa}
\end{equation}%
where the function $ang$ represents the phase angle of a complex value. In
the case of $\phi \ll 1$ and $\phi /\varepsilon \ll 1$, the transformed
polarization state is approximately equivalent to $( |H\rangle +\varepsilon
e^{\frac{i\phi }{2\varepsilon }}|V\rangle) /\sqrt{1+\varepsilon ^{2}}$, from
which one can clearly find out that the relative phase shift is amplified.
After passing through a polarizing beam splitter in the interferometer, this
state can be written as $( |H\rangle|0\rangle +\varepsilon e^{\frac{i\phi }{%
2\varepsilon }}|V\rangle|1\rangle) /\sqrt{1+\varepsilon ^{2}}$, where $%
\left\{ |0\rangle ,|1\rangle \right\} $ represents the path degree of
freedom in the interferometer.

Further, in order to suppress the phase uncertainty induced by the systematic phase errors and thus improve
the phase resolution of interferometer with such transformed polarization
state, it is still required to attenuate the intensity of $|H\rangle $
component down to the same level as $|V\rangle $ component. After the
attenuation, the final qubit state is changed to $\frac{1}{\sqrt{2}}%
(|H\rangle +e^{\frac{i\phi }{2\varepsilon }}|V\rangle ),$ with a successful
probability of $\thicksim 2\varepsilon ^{2}$. We note that attenuating the
intensity of $|H\rangle $ component is equivalent to project the path state
to $|f\rangle =\left( \varepsilon |0\rangle +|1\rangle \right) /\sqrt{%
1+\varepsilon ^{2}}.$ Due to the effect of phase shift amplification, the
phase uncertainty of the interferometer is thus improved to (see Eq. (A16) in Appendix A)
\begin{equation}
\Delta \phi =2\sqrt{\varepsilon ^{2}\rho ^{2}+\frac{1}{2N}}.  \label{pu2}
\end{equation}%
Although the contribution of statistical phase errors is multiplied by $\sqrt{%
2}$, the contribution of systematic phase errors is drastically suppressed by
a factor of $2\varepsilon $. Therefore, with the help of weak measurement,
the phase resolution of interferometer can be improved by a factor of $\frac{%
1}{2\varepsilon }$, when $\rho \gg \frac{1}{\sqrt{N}}$.

From the perspective of weak measurement, the path degree of freedom can be
treated as a \textit{system}, and the polarization degree of freedom can be
treated as a \textit{qubit meter}~\cite{Wu09}. The coupling of the \emph{%
system} and the \emph{meter} is described by%
\begin{equation}
\frac{1}{\sqrt{2(1+\varepsilon ^{2})}}e^{\frac{1}{4}i\pi \sigma _{y}\otimes
\left( \sigma _{z}-I\right) }\left( |L\rangle +|R\rangle \right) \otimes
\left( |H\rangle +\varepsilon e^{i\phi ^{\prime }}|V\rangle \right) ,
\end{equation}%
where $|L\rangle $ and $|R\rangle $ are the eigenstates of Pauli matrix $%
\sigma _{y}$, i.e., $\frac{1}{\sqrt{2}}\left( |0\rangle \pm i|1\rangle
\right) $. This description can be rewritten as
\begin{equation}
\frac{1}{\sqrt{2}}e^{-\frac{1}{2}i\theta \sigma _{y}\otimes \sigma _{\vec{r}%
}}\left( |L\rangle +|R\rangle \right) \otimes |H\rangle ,  \label{weak int}
\end{equation}%
where $\theta $ is the azimuth angle of the Bloch vector of the meter state
and $\sigma _{\vec{r}}=\cos \phi ^{\prime }\sigma _{x}+\sin \phi ^{\prime
}\sigma _{y}$. When $\phi \ll 1$ and $\phi /\varepsilon \ll 1,$ $\theta $ is
approximately equal to $2\varepsilon $, which shows the weak interaction
between the \emph{system} and the \emph{meter}. When the \emph{system} is
post-selected to the final state $|f\rangle $, the \emph{meter} state can be
read out. From Eq.~\ref{weak int}, with weak interaction approximation $%
e^{-i\varepsilon \sigma _{y}\otimes \sigma _{\vec{r}}}\approx 1-i\varepsilon
\sigma _{y}\otimes \sigma _{\vec{r}}$, the final \emph{meter} state is%
\begin{eqnarray}
&&\left[ \left\langle f\right\vert \left( 1-i\varepsilon \sigma _{y}\otimes
\sigma _{\vec{r}}\right) \Vert i\rangle \right] |H\rangle  \notag \\
&\approx &\varepsilon (|H\rangle +e^{i\phi ^{\prime }}|V\rangle ).
\label{post selection}
\end{eqnarray}%
Since $\left\langle f|i\right\rangle $ is at the same order of $\varepsilon $%
, Eq. \ref{post selection} cannot be transformed to $exp[\frac{%
-i\varepsilon \left\langle f|\sigma _{y}|i\right\rangle }{\left\langle
f|i\right\rangle }\sigma _{\vec{r}}]|H\rangle $, as the standard formula of
weak measurements \cite{Aharonov88}. Therefore, in our scheme there dose not exist a generally
defined weak value. More importantly, the successful probability
in our scheme ($\thicksim 2\varepsilon ^{2}$) is twice as in the scenario with standard weak measurements given the same
amplification factor (see Appendix B for details).

\begin{figure}[tb]
\centering
\includegraphics[width=7 cm]{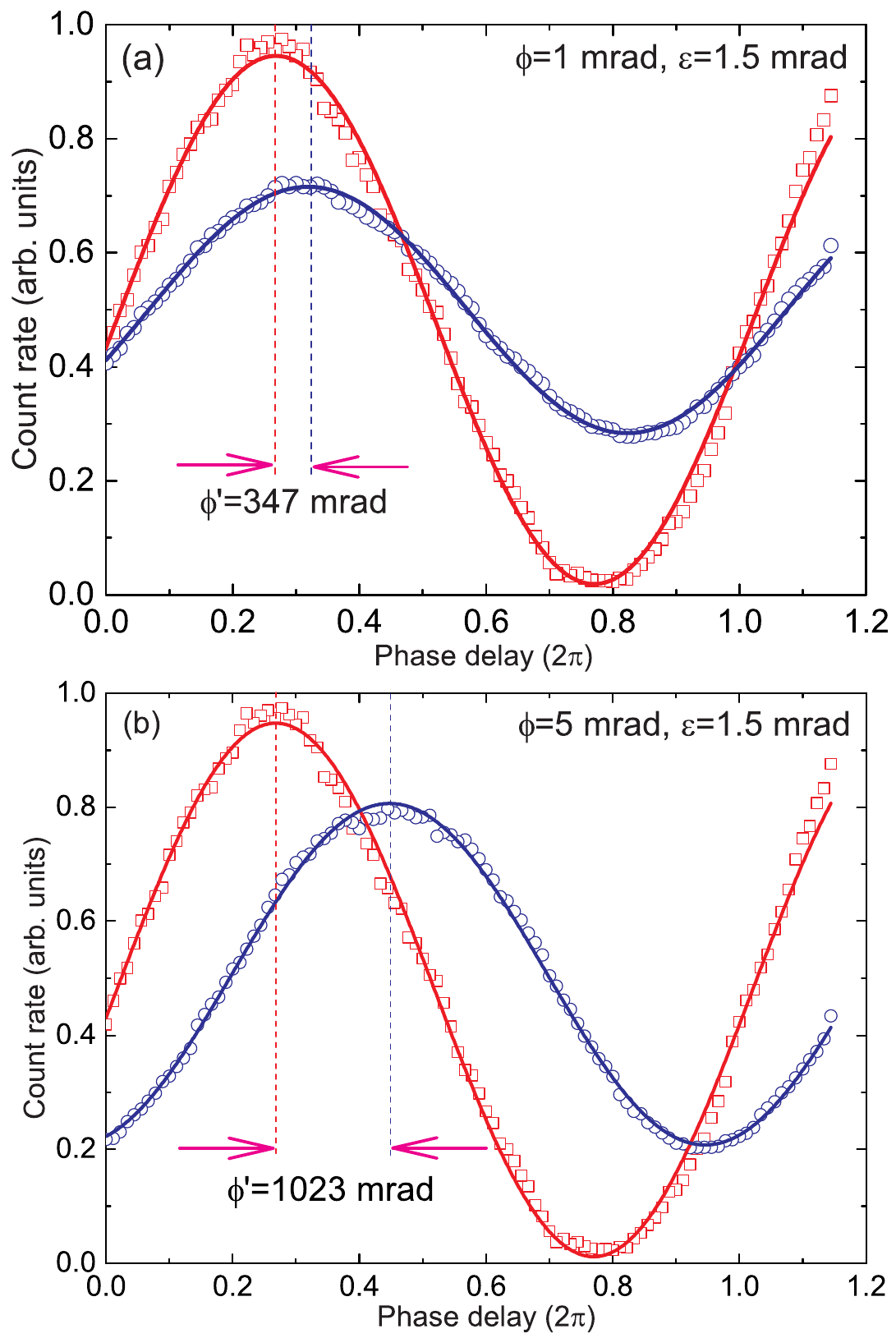}
\caption{Measured oscillation patterns without (squares) and with (circles) weak measurement. $\protect\varepsilon$ is set as 1.5 mrad, and $%
\protect\phi$ is set as 1 mrad (a) and 5 mrad (b), respectively. By
sinusoidally fitting the oscillation patterns in two cases (solid lines),
the final phase shifts $\protect\phi ^{\prime }$ with weak measurement can
be calculated, which are much larger than the initial phase shifts $\protect%
\phi$. Such phase amplification highly depends on the parameters of $\protect%
\phi$ and $\protect\varepsilon.$ }
\label{fig2}
\end{figure}

\section{experiment}

We then perform an experiment to demonstrate the phase shift amplification
based on our weak measurement scheme. The experimental setup is shown in
Fig.~\ref{fig1}. A continuous wave laser with a center wavelength of 785 nm
is attenuated down to single-photon level with a set of attenuators (ATT),
and the polarization state of incident photons is prepared to $\frac{1}{%
\sqrt{2}}(|H\rangle +|V\rangle )$ by a polarizer with PER of $\sim $ $%
10^{4}:1$. The initial phase shift $\phi $ between $|H\rangle $ and $%
|V\rangle $ to be measured is produced by a Soleil-Babinet Compensator (SBC)
with a high-precision phase modulation step of $0.26$ mrad. The SBC is a
zero-order retarder that can be adjusted continuously.

The weak measurement part consists of two components, a half-wave plate
(HWP1) that transforms the initial state $\frac{1}{\sqrt{2}}(|H\rangle
+e^{i\phi }|V\rangle )$ to $(|H\rangle +\varepsilon e^{\frac{i\phi}{%
2\varepsilon }}|V\rangle )/\sqrt{1+\varepsilon ^{2}}$, and a Sagnac-like
interferometer. $|H\rangle $ and $|V\rangle $ photons are separated into two
different paths of the interferometer. Two polarizers are inserted into the
transmitted and reflected paths to further improve the polarization
visibility, respectively, and two attenuators are used to independently
adjust the intensities of $|H\rangle $ and $|V\rangle $ photons. At the
output port of the interferometer, polarization projection measurements is
performed using a quarter-wave plate (QWP) set at $45^{\circ }$, HWP2, a
Glan-Laser Calcite Polarizer (GL) with high PER of $\sim 10^5:1$, and a
Silicon avalanche photodiode (APD). By tuning the angle of HWP2 to modulate
the phase delay between $|H\rangle $ and $|V\rangle $, oscillation patterns
are scanned, and thus the phase shifts between the oscillation patterns can
be directly measured.

In the experiment, all the HWPs and polarizers are mounted on stepper
electric motors, whose minimal rotation angles can reach as low as $0.52$
mrad. Fig.~\ref{fig2} shows the measured polarization oscillation patterns
by tuning the angle of HWP2 with a phase delay step of $1^{\circ }$, and the
phase shifts between the oscillation patterns are calculated using the
sinusoidal fitting curves. The polarization visibility degradation as shown
in Fig.~\ref{fig2} is mainly due to the operations of intensity balance.
With $\varepsilon=$ 1.5 mrad, the final phase shifts $\phi ^{\prime }$ reach
347 mrad, 1023 mrad under the settings of $\phi=$ 1 mrad, $\phi=$ 5 mrad,
respectively, which clearly exhibits the phase amplification effect due to
weak measurement. The phase amplification gain highly depends on the
parameters of $\phi$ and $\varepsilon $. We note that without phase
amplification the original phase resolution of the interferometer in the
experiment using standard linear optics devices is around $10$ mrad, which
is limited by the PER of GL.

\begin{figure}[tb]
\centering
\includegraphics[width=7 cm]{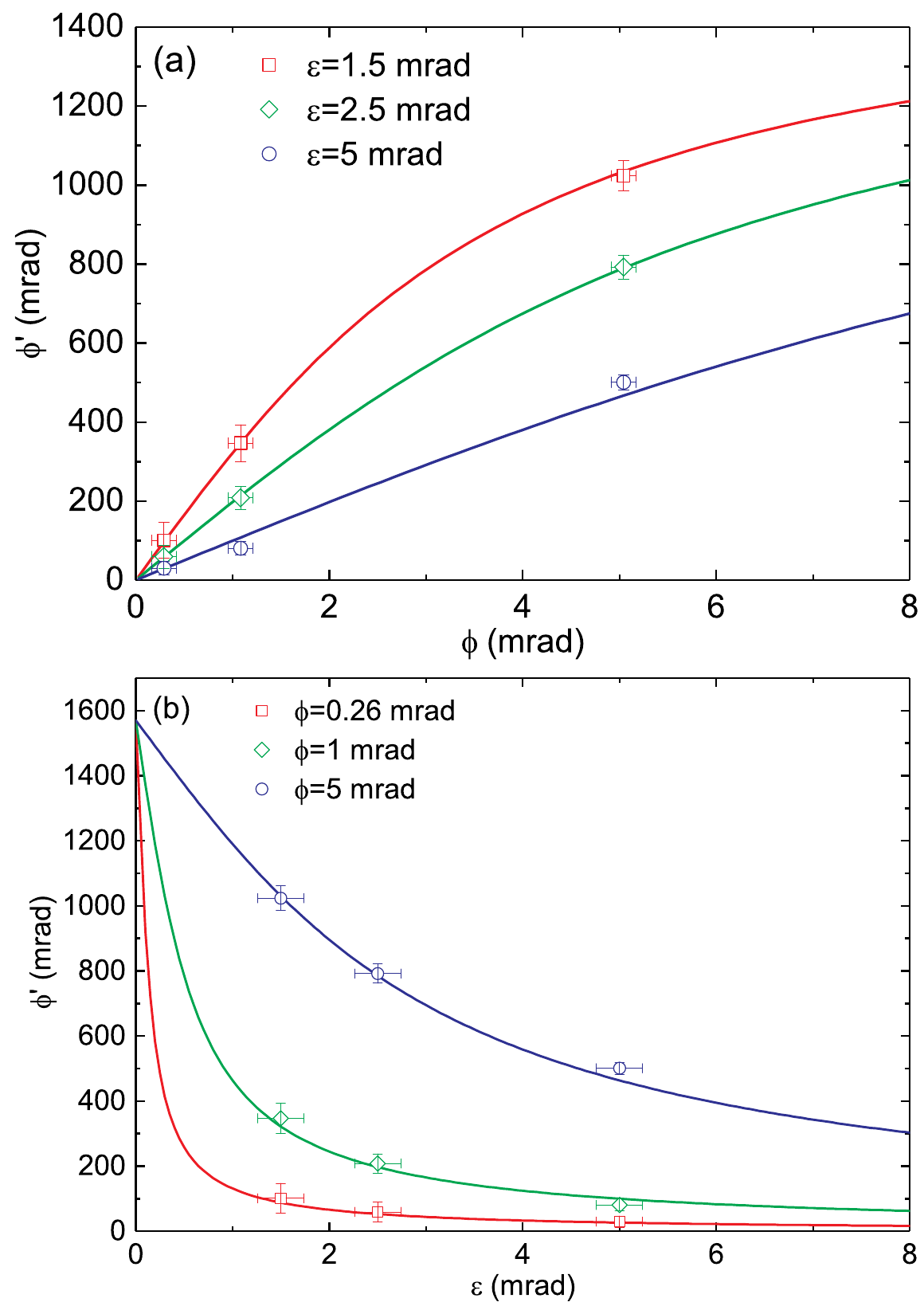}
\caption{The plots of the final phase shifts $\protect\phi ^{\prime }$ as a
function of the initial phase shift $\protect\phi $ (a) and the rotation
angle $\protect\varepsilon $ (b). Points and solid lines represent
experimental values and theoretical fits, respectively.}
\label{fig3}
\end{figure}

Further, we investigate the quantitative relationships of three parameters,
i.e., initial phase shift $\phi $, rotation angle of deviation $\varepsilon $
and final phase shift $\phi^{\prime }$. The measured results are plotted in
Fig.~\ref{fig3}. Fig.~\ref{fig3}(a) shows the relationship between $\phi $
and $\phi ^{\prime }$ with different settings of $\varepsilon $. The
theoretical fitting curves are calculated according to Eq. \ref{pa}. $\phi
^{\prime }$ increases with the increase of $\phi $, and with a fixed value
of $\phi $ rotation angles that are closer to $\pi /4$ result in larger
phase shift amplification. Fig.~\ref{fig3} (b) shows the relationship
between $\phi ^{\prime }$ and $\varepsilon $ with different settings of $%
\phi $. $\phi ^{\prime }$ decreases with the increase of $\varepsilon $, and
with a fixed value of $\varepsilon $ larger initial phase shifts result in
larger final phase shifts. Particularly, the slopes of phase amplification
are steep in the regime of small $\varepsilon $, but become flat when $%
\varepsilon >$ $5$ mrad. From Fig.~\ref{fig3}, one can find out that using
weak measurement the phase resolution can be improved to $0.26$ mrad at
least for the interferometer, since in the experiment the initial phase
shift cannot be generated smaller than $0.26$ mrad due to the limit of SBC
adjustment. This result is already significantly better than the original phase
resolution of the interferometer without weak measurement. Nevertheless, in
principle smaller values of $\phi $ can be resolved with our method, since
the phase amplification gain is considerably high in the regime of small
angles of $\varepsilon .$ For instance, with $\phi =0.26$ mrad and $%
\varepsilon =1.5$ mrad, the final phase shift $\phi ^{\prime }$ reaches $101$
mrad, corresponding to a phase amplification gain of more than $388$.

\section{discussion}

Our weak measurement scheme is compatible with the conventional approaches
for phase resolution improvement in optical interferometry, especially the
technique of signal proceeding, which makes it possible to perform extremely
small phase measurement. For instance, an original phase shift measurement
of $10^{-7}$ rad with the lock-in amplifier in an optical interferometer was
reported~\cite{Matsuda09}. By using the phase shifter device in that
experiment instead of the SBC in our experiment and performing signal
proceeding as usual, the phase amplification effect in our scheme may yield
further enhancement in the phase sensitivity by 2$\sim$3 orders of
magnitudes so that to bring the measurement precision of phase shift to the
level of $10^{-9}$ rad.

Furthermore, our weak measurement scheme can also be adapted for $N00N$
states to achieve better phase resolution. For instance, one can prepare $%
N00N$ states $\frac{1}{\sqrt{2}}(|H\rangle ^{\otimes n}+|V\rangle ^{\otimes
n})$ as input to the SBC instead of single-photon states, after projecting $%
n-1$ photonic qubits to $\frac{1}{\sqrt{2}}(|H\rangle \pm |V\rangle )$ as
implemented in the previous $N00N$ state experiments~\cite%
{Walther04,Leibfried04,Gao10} and performing the same unitary rotation and
post-selection operation on the $n$th qubit as required in our weak
measurement scheme, the final phase shift can be amplified to $\frac{n\phi }{%
2\varepsilon }$. The total phase uncertainty in such case is (see Eq. (A19) in Appendix A)
\begin{equation}
\Delta \phi =2\sqrt{\frac{\varepsilon ^{2}\rho ^{2}}{n^2}+\frac{1}{2nN}}.  \label{pu3}
\end{equation}%
Compared to Eq.~\ref{pu1}, one can conclude that weak measurement scheme
with $N00N$ states can not only significantly decrease the systematic phase uncertainty
but also suppress the statistical phase uncertainty and particularly beat
the standard quantum limit when $n>2$.

\section{summary}

In summary, we have proposed and experimentally demonstrated direct phase
shift amplification in optical interferometry with weak measurement, which
drastically improves the phase shift amplification with a factor of $388$.
Since phase measurement is a fundamental metrology tool, such weak
measurement scheme can provide a new approach for the precision measurement
applications of various physical quantities including observation of weak
cross-Kerr nonlinearity~\cite{Feizpour11,MTS17} and test of general relativity with quantum
interference~\cite{Zych11}.

\section*{acknowledgments}

The authors thank Y.-A. Chen for insightful discussions. This work has been supported by the National Key R\&D Program of China under Grant No.~2017YFA0304004, the National Natural Science Foundation of China under Grant No.~91336214, No.~11574297, and No.~11374287, and the Chinese Academy of Sciences.

\setcounter{equation}{0}
\renewcommand\theequation{A\arabic{equation}}

\begin{appendix}

\section*{appendix a: Phase uncertainty in interferometry with systematic error}

Considering a phase measurement in a realistic interferometer, without loss
of generality it is assumed that the systematic phase error $\varphi $ is a
zero-mean value following a Gaussian distribution,
\begin{equation}
D\left( \varphi \right) \varpropto e^{-\frac{\varphi ^{2}}{2\rho ^{2}}},
\end{equation}%
where $\rho $ is the standard derivation. The detection probability at one
output port of the interferometer is
\begin{equation}
p=\frac{1}{2}(1+\cos (\phi +\varphi )),  \label{p}
\end{equation}%
where $\phi $ is the phase to be measured. Given a certain fixed phase error
$\varphi $, the sample size interval is $N_{\varphi }=N{\cdot }D\left(
\varphi \right) \delta \varphi $. For a large total sample size $N$, the
variance of estimated probability $p_{\varphi }$ in each sample interval can
be calculated as
\begin{equation}
\left( \Delta p_{\varphi }\right) ^{2}=\frac{p_{\varphi }\left( 1-p_{\varphi
}\right) }{N_{\varphi }}=\frac{\sin ^{2}\left( \phi +\varphi \right) }{%
4N_{\varphi }},  \label{1}
\end{equation}%
where $\left\langle p_{\varphi }\right\rangle =p$.

From Eq.~\ref{1}, the mean values of $p$ and $p^2$ during the whole
sampling process can be further calculated as
\begin{eqnarray}
\left\langle p\right\rangle &=&\frac{\sum N_{\varphi }\left\langle
p_{\varphi }\right\rangle }{N}\approx \int D\left( \varphi \right)
\left\langle p_{\varphi }\right\rangle d\varphi  \notag \\
&=&\frac{1}{2}+\frac{1}{2}\cos \phi \int D\left( \varphi \right) \cos
\varphi d\varphi,  \label{2}
\end{eqnarray}%
\begin{eqnarray}
\left\langle p^{2}\right\rangle &=&\frac{\sum N_{\varphi }\left\langle
p_{\varphi }^{2}\right\rangle }{N}=\frac{\sum N_{\varphi }\left[ \left(
\Delta p_{\varphi }\right) ^{2}+\left\langle p_{\varphi }\right\rangle %
\right] }{N}  \notag \\
&\approx &\frac{1}{4N}\int D\left( \varphi \right) \sin ^{2}\left( \phi
+\varphi \right) d\varphi  \notag \\
&+&\frac{1}{4}\int D\left( \varphi \right) \left[ 1+\cos \left( \phi
+\varphi \right) \right] ^{2}d\varphi,  \label{3}
\end{eqnarray}%
respectively. Combining Eq.~\ref{2} and Eq.~\ref{3}, the variance of $p$
is
\begin{eqnarray}
\left( \Delta p\right) ^{2} &=&\left\langle p^{2}\right\rangle -\left\langle
p\right\rangle ^{2}  \notag \\
&=&\frac{1}{4}\int \frac{1}{N}\left( \sin ^{2}\phi \cos ^{2}\varphi +\cos
^{2}\phi \sin ^{2}\varphi \right) D\left( \varphi \right) d\varphi  \notag \\
&+&\left( \cos ^{2}\phi \cos ^{2}\varphi +\sin ^{2}\phi \sin ^{2}\varphi
\right) D\left( \varphi \right) d\varphi  \notag \\
&-&\frac{1}{4}\cos ^{2}\phi \left[ \int \cos \varphi D\left( \varphi \right)
d\varphi \right] ^{2}.  \label{4}
\end{eqnarray}%
Since $\rho \ll 1$ and $\varphi $ is a small quantity in the integrals,
therefore, one can consider only the terms of $\varphi $ and $\varphi ^{2}$
with the following approximations
\begin{eqnarray}
&\int&\sin ^{2}\varphi D\left( \varphi \right) d\varphi \approx \rho ^{2},
\notag \\
&\int&\cos ^{2}\varphi D\left( \varphi \right) d\varphi \approx 1-\rho ^{2},
\notag \\
&\int&\cos \varphi D\left( \varphi \right) d\varphi \approx 1-\frac{\rho
^{2}}{2}.  \label{approx}
\end{eqnarray}%
According to Eq.~\ref{approx}, Eq.~\ref{4} can be simplified with
approximations as
\begin{equation}
\left( \Delta p\right) ^{2}=\frac{\sin ^{2}\phi }{4}\left( \rho ^{2}+\frac{1%
}{N}+\frac{\cos ^{2}\phi\rho ^{2}}{N\sin ^{2}\phi }\right).  \label{deltap}
\end{equation}%
Combining Eq.~\ref{p} and Eq.~\ref{deltap}, the phase uncertainty is
calculated as
\begin{eqnarray}
\Delta \phi &=&\left\vert \frac{d\phi }{dp}\cdot \Delta p\right\vert  \notag
\\
&=&\sqrt{\rho ^{2}+\frac{1}{N}+\frac{\rho ^{2}}{N\sin ^{2}\phi }}.
\label{deltaphi}
\end{eqnarray}

Due to the existence of phase errors, the theoretically maximum estimated
probability is $1-\frac{1}{N}$. From Eq.~\ref{p}, one can calculate the
minimum measurable phase by $\sin^{2}\phi =\frac{4}{N}$. Therefore, the
phase uncertainty in a realistic interferometer can be roughly estimated as
\begin{equation}
\Delta \phi =\sqrt{\rho ^{2}+\frac{1}{N}}.
\end{equation}

Further, one can calculate the phase uncertainty in the following cases,
i.e., using $N00N$ state, using single-qubit state with weak measurement and
using $N00N$ state with weak measurement.

\emph{$N00N$ state}. With a sample size $m$ satisfying $n\cdot m=N$, the
detection probability at one output port of an interferometer is
\begin{equation}
p^{(N)} = \frac{1}{2}\left( 1+\cos \left( n\phi +\varphi \right) \right).
\end{equation}
The mean value and the variance of estimated probability in each sampling
interval are
\begin{eqnarray}
\left\langle p_{\varphi}^{(N)}\right\rangle &=& p^{(N)} ,  \notag \\
\left( \Delta p_{\varphi}^{(N)}\right) ^{2} &=&\frac{n\sin ^{2}\left( n\phi
+\varphi \right) }{4N_{\varphi }},
\end{eqnarray}%
respectively.\ Therefore, the phase uncertainty using $N00N$ states in a
realistic interferometer is%
\begin{equation}
\Delta \phi =\sqrt{\frac{\rho ^{2}}{n^2}+\frac{1}{nN}}.
\end{equation}

\emph{Single-qubit state with weak measurement.} As described in the text,
in such case the detection probability at one output port of an
interferometer is
\begin{equation}
p_{W} = \frac{1}{2}\left( 1+\cos \left(\frac{\phi} {2\varepsilon} +\varphi
\right) \right),
\end{equation}
with a sample size of $2\varepsilon ^{2}N$. Then, the mean value and the
variance of estimated probability in each sampling interval are
\begin{eqnarray}
\left\langle p_{\varphi W}\right\rangle &=& p_{W},  \notag \\
\left( \Delta p_{\varphi W}\right) ^{2} &=&\frac{\sin ^{2}\left( \frac{\phi
}{2\varepsilon }+\varphi \right) }{2\varepsilon ^{2}N_{\varphi }},
\end{eqnarray}%
respectively. Therefore, the phase uncertainty using single-qubit states
with weak measurement in a realistic interferometer is%
\begin{equation}
\Delta \phi =2\sqrt{\varepsilon ^{2}\rho ^{2}+\frac{1}{2N}}.
\end{equation}

\emph{$N00N$ state with weak measurement.} As described in the text, in such
case the detection probability at one output port of an interferometer is
\begin{equation}
p_{W}^{(N)}=\frac{1}{2}\left( 1+\cos \left( \frac{n\phi }{2\varepsilon }%
+\varphi \right) \right) ,
\end{equation}%
with a sample size of $\frac{2\varepsilon ^{2}N}{n}$. Then, the mean value
and the variance of estimated probability in each sampling interval are
\begin{eqnarray}
\left\langle p_{\varphi W}^{(N)}\right\rangle &=&p_{W}^{(N)},  \notag \\
\left( \Delta p_{\varphi W}^{(N)}\right) ^{2} &=&\frac{n\sin ^{2}\left(
\frac{n\phi }{2\varepsilon }+\varphi \right) }{2\varepsilon ^{2}N_{\varphi }%
},
\end{eqnarray}%
respectively. Therefore, the phase uncertainty using $N00N$ states with weak
measurement in a realistic interferometer is%
\begin{equation}
\Delta \phi =2\sqrt{\frac{\varepsilon ^{2}\rho ^{2}}{n^2}+\frac{1}{2nN}}.
\end{equation}

\begin{figure}[t]
\centering\includegraphics [width=7.5 cm]{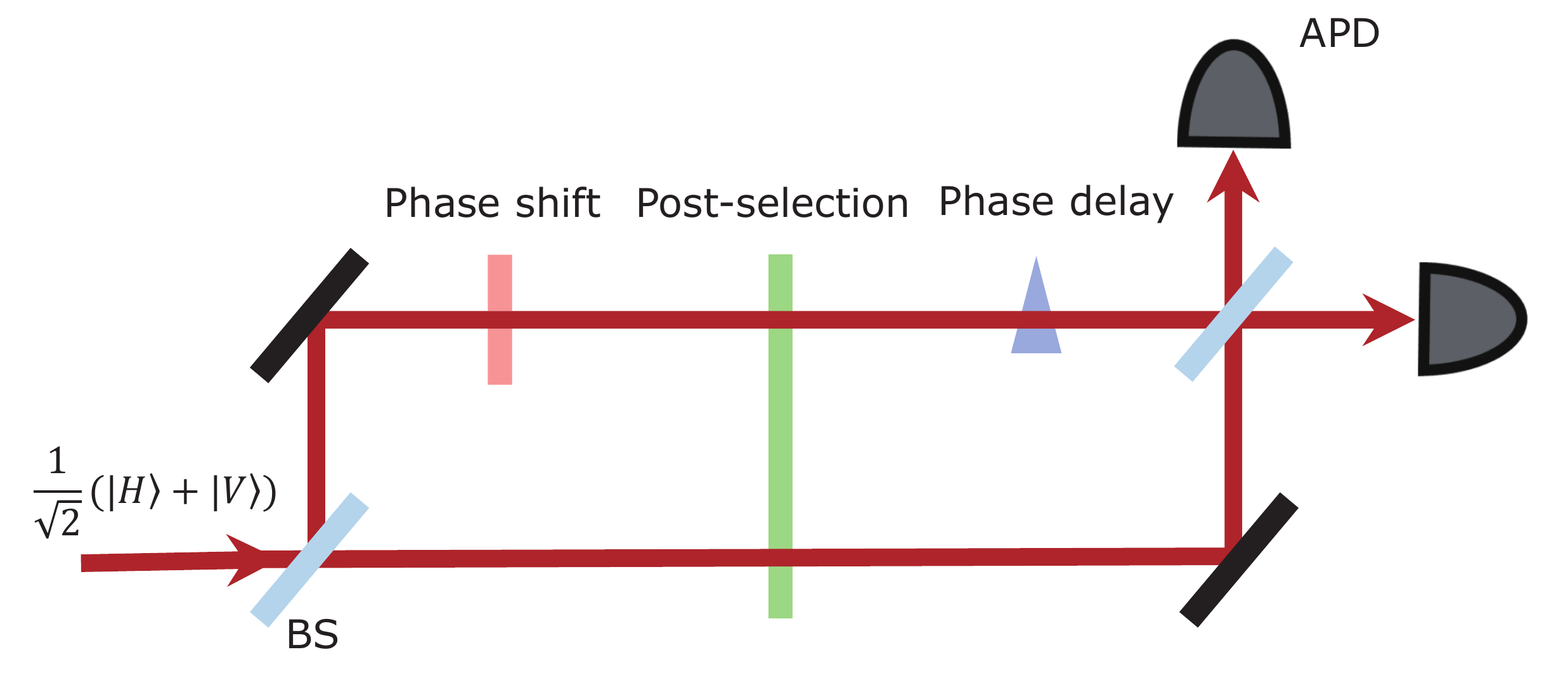}
\caption{Phase amplification scheme with standards weak measurement.
In the upper path of the interferometer, a tiny phase shift $\protect\phi $
is induced between $|H\rangle $ and $|V\rangle $. The post-selection is set
to the almost orthogonal state of the incident polarization states. The tunable
phase delay in the path is used to measure the oscillation patterns. }
\label{fig4}
\end{figure}

\section*{appendix b: Phase amplification scheme with standard weak measurements}

In this scheme, the polarization degree of freedom can be treated as a
\textit{system}, and the path degree of freedom can be treated as a \textit{%
qubit meter}.

As shown in Fig.~\ref{fig4}, the initial \textit{system} and \textit{meter}
state are $|i\rangle =\frac{1}{\sqrt{2}}(|H\rangle +|V\rangle )$ and $\frac{1%
}{\sqrt{2}}(|0\rangle +|1\rangle )$, respectively, where the $|0\rangle $
and $|1\rangle $ are labels of the upper and lower paths of Mach-Zehnder
interferometer. Given a tiny phase shift $\phi $ between $|H\rangle $ and $%
|V\rangle $ in the path $|0\rangle $ to be measured, the \textit{system-meter%
} composite state is
\begin{equation}
\frac{1}{2}[|H\rangle (|0\rangle +e^{i\phi /2}|1\rangle )+|V\rangle
(|0\rangle +e^{-i\phi /2}|1\rangle )].
\end{equation}Therefore, the coupling of the \textit{system} and \textit{meter }can be
written as
\begin{equation}
\frac{1}{2}e^{-\frac{i\phi }{4}\sigma _{z}\otimes \sigma _{z}}\left(
|H\rangle +|V\rangle \right) \otimes \left( |0\rangle +|1\rangle \right),
\end{equation}%
where $\phi \ll 1$ shows the weak interaction. Then, the \textit{system}
state is post-selected to the almost orthogonal final state $|f\rangle =\sin
(\pi /4+\varepsilon )|H\rangle -\cos (\pi /4+\varepsilon )|V\rangle $ ($%
\varepsilon \ll 1$) in both paths. When $\phi /\varepsilon \ll 1,$ the final
\textit{meter} state is approximately
\begin{eqnarray}
&&\frac{1}{\sqrt{2}}\left\langle f\right\vert \left( 1-\frac{i\phi }{4}%
\sigma _{z}\otimes \sigma _{z}\right) |i\rangle \otimes \left( |0\rangle
+|1\rangle \right) \notag \\
&\approx &\frac{\left\langle f|i\right\rangle }{\sqrt{2}}e^{\frac{-i\phi
\left\langle f\right\vert \sigma _{z}\left\vert i\right\rangle }{%
4\left\langle f|i\right\rangle }\sigma _{z}}\left( |0\rangle +|1\rangle
\right) \notag \\
&=&\frac{\left\langle f|i\right\rangle }{\sqrt{2}}\left( |0\rangle +e^{\frac{%
i\phi }{2\varepsilon }}|1\rangle \right) ,
\end{eqnarray}%
where the weak value is $A_{W}\equiv\frac{\left\langle f\right\vert \sigma
_{z}\left\vert i\right\rangle }{\left\langle f|i\right\rangle }\approx \frac{%
1}{\varepsilon }$ and the successful probability is $\left\vert \left\langle
f|i\right\rangle \right\vert ^{2}\approx $ $\varepsilon ^{2}.$

\end{appendix}



\begin{thebibliography}{99}
\bibitem{Giovannetti04} V. Giovannetti, S. Lloyd, and L. Maccone,
Science \textbf{306}, 1330-1336 (2004).

\bibitem{Giovannetti11} V. Giovannetti, S.Lloyd, and L. Maccone,
Nat. Photon. \textbf{5}, 222-229 (2011).

\bibitem{Boto00} A. Boto, P. Kok, D. S. Abrams, S. L. Braunstein, C. P. Williams, and J. P. Dowling,
Phys. Rev. Lett. \textbf{85}, 2733-2736 (2000).

\bibitem{Holland93} M. J. Holland and K. Burnett,
Phys. Rev. Lett. \textbf{71}, 1355-1358 (1993).

\bibitem{Caves81} C. M. Caves,
Phys. Rev. D \textbf{23}, 1693-1708 (1981).

\bibitem{Wang16} X.-L. Wang, L.-K. Chen, W. Li, H.-L. Huang, C. Liu, C. Chen, Y.-H. Luo, Z.-E. Su, D. Wu, Z.-D. Li, H. Lu, Y. Hu, X. Jiang, C.-Z. Peng, L. Li, N.-L. Liu, Y.-A. Chen, C.-Y. Lu, and J.-W. Pan,
Phys. Rev. Lett. \textbf{117}, 210502 (2016).

\bibitem{Chen17} L.-K. Chen, Z.-D. Li, X.-C. Yao, M. Huang, W. Li, H. Lu, X. Yuan, Y.-B. Zhang, X. Jiang, C.-Z. Peng, L. Li, N.-L. Liu, X. Ma, C.-Y. Lu, Y.-A. Chen, and J.-W. Pan,
Optica \textbf{4}, 77-83 (2017).

\bibitem{Aharonov88} Y. Aharonov, D. Z. Albert, and L. Vaidman,
Phys. Rev. Lett. \textbf{60}, 1351-1354 (1988).

\bibitem{Aharonov90} Y. Aharonov and L. Vaidman,
Phys. Rev. A \textbf{41}, 11-20 (1990).

\bibitem{Feizpour11} A. Feizpour, X. Xing, and A. M. Steinberg,
Phys. Rev. Lett. \textbf{107}, 133603 (2011).

\bibitem{Zilberberg11} O. Zilberberg, A. Romito, and Y. Gefen,
Phys. Rev. Lett. \textbf{106}, 080405 (2011).

\bibitem{Wu12} S. Wu and M. Zukowski,
Phys. Rev. Lett. \textbf{108}, 080403 (2012).

\bibitem{Strubi13} G. Str\"{u}bi and C. Bruder,
Phys. Rev. Lett. \textbf{110}, 083605 (2013).

\bibitem{Pang14} S. Pang, J. Dressel, and T. A. Brun,
Phys. Rev. Lett. \textbf{113}, 030401 (2014).

\bibitem{Hosten08} O. Hosten and P. Kwiat,
Science \textbf{319}, 787-790 (2008).

\bibitem{Dixon09} P. B. Dixon, D. J. Starling, A. N. Jordan, and J. C.
Howell,
Phys. Rev. Lett. \textbf{102}, 173601 (2009).

\bibitem{Gorodetski12} Y. Gorodetski, K. Y. Bliokh, B. Stein, C. Genet, N.
Shitrit, V. Kleiner, E. Hasman, and T. W. Ebbesen,
Phys. Rev. Lett. \textbf{109}, 013901 (2012).

\bibitem{Magana14} O. S. Maga\~{n}a-Loaiza, M. Mirhosseini, B. Rodenburg,
and R. W. Boyd,
Phys. Rev. Lett. \textbf{112}, 200401 (2014).

\bibitem{Brunner10} N. Brunner and C. Simon,
Phys. Rev. Lett. \textbf{105}, 010405 (2010).

\bibitem{Xu13} X.-Y. Xu, Y. Kedem, K. Sun, L. Vaidman, C.-F. Li, and G.-C. Guo,
Phys. Rev. Lett. \textbf{111}, 033604 (2013).

\bibitem{SS14}
L. J. Salazar-Serrano, D. Janner, N. Brunner, V. Pruneri, and J. P. Torres,
Phys. Rev. A \textbf{89}, 012126 (2014).

\bibitem{Xiaodong17}
X. Qiu, L. Xie, X. Liu, L. Luo, Z. Li, Z. Zhang, and J. Du,
Appl. Phys. Lett. \textbf{110}, 071105 (2017).

\bibitem{Wu09} S. Wu and K. M{\o }lmer,
Phys. Lett. A \textbf{374}, 34-39 (2009).

\bibitem{Matsuda09} N. Matsuda, R. Shimizu, Y. Mitsumori, H. Kosaka, and K.
Edamatsu,
Nat. Photon. \textbf{3}, 95-98 (2009).

\bibitem{Walther04} P. Walther, J.-W. Pan, M. Aspelmeyer, R. Ursin, S.
Gasparoni, and A. Zeilinger,
Nature \textbf{429}, 158-161 (2004).

\bibitem{Leibfried04} D. Leibfried, M. D. Barrett, T. Schaetz, J. Britton,
J. Chiaverini, W. M. Itano, J. D. Jost, C. Langer, and D. J. Wineland,
Science \textbf{304}, 1476-1478 (2004).

\bibitem{Gao10} W.-B. Gao, C.-Y. Lu, X.-C. Yao, P. Xu, O. G\"{u}hne, A.
Goebel, Y.-A. Chen, C.-Z. Peng, Z.-B. Chen, and J.-W. Pan,
Nat. Phys. \textbf{6}, 331-335 (2010).

\bibitem{MTS17}
F. Matsuoka, A. Tomita, and Y. Shikano,
Quantum Stud.: Math. Found. \textbf{4}, 159-169 (2017).

\bibitem{Zych11} M. Zych, F. Costa, I. Pikovski, and C. Brukner,
Nat. Commun. \textbf{2}, 505 (2011).
\end{thebibliography}
\end{document}